\documentstyle[pra,aps]{revtex}
          \begin{document}
	   \draft
          \title{Quantum Robots Plus Environments} 
          \author{Paul Benioff\\
           Physics Division, Argonne National Laboratory \\
           Argonne, IL 60439 \\
           e-mail: pbenioff@anl.gov}
           \date{\today}

          \maketitle
          \begin{abstract}  
A quantum robot is a mobile quantum system, including an on board
quantum computer and needed ancillary systems, that interacts
with an environment of quantum systems.  Quantum robots carry out
tasks whose goals include making specified changes in the state
of the environment or carrying out measurements on the
environment.  The environments considered so far, oracles,
data bases, and quantum registers, are seen to be special cases of
environments considered here. It is also seen that a
quantum robot should include a quantum computer and cannot be
simply a multistate head.

A model of quantum robots and their interactions is discussed in
which each task, as a sequence of alternating computation and
action phases,is described by a unitary single time step operator
${\cal T}={\cal T}_{a}+{\cal T}_{c}$ (discrete space and time are assumed). 
The overall system dynamics is described as a sum over paths of
completed computation (${\cal T}_{c}$) and action (${\cal T}_{a}$) phases.  A
simple example of a task, measuring the distance between the
quantum robot and a particle on a 1D lattice with quantum phase
path dispersion present, is analyzed. A decision diagram for the
task is presented and analyzed.
 \end{abstract}

\section{Introduction}
 
Quantum computers are of much interest due to their increased
power over classical computers in solving certain problems
\cite{Shor,Grover}.  Most studies of quantum computers consider
them as stand alone systems operating in isolation from external
systems as an environment. So far work on quantum computers which
includes interactions with environments is limited mainly to
noise effects, data base searching, and quantum oracle computing. 
The former considers environmental interactions as a source of
noise and errors. This has stimulated the development of quantum
error correcting codes to minimize this effect \cite{Laf,Zur}
Other methods rely on the use of properties of systems with
relatively long decoherence times \cite{Div,Ger}.  Quantum oracle
computing has been discussed but not developed extensively
\cite{Bennett}.  Data base searching has been much discussed
recently \cite{Grover}.

Here the emphasis is on quantum computers and their interactions
with the environment in general.  The interest is in quantum
computers along with ancillary systems that can move in and
interact with an environment of quantum systems. These are the
defining characteristics of quantum robots.

Quantum robots are of interest from a foundational viewpoint \cite{Ben1}. If
quantum mechanics is universally valid, then the systems that carry out
theoretical calculations (computers) and physical experiments to test theoretical
predictions (robots) must be described within quantum mechanics, i. e. as
quantum computers and quantum robots. It follows that the systems that test
the validity of quantum mechanics must be described by the same theory they
are testing.  Quantum mechanics must describe its own validity to the
maximum extent possible \cite{PerZur}.

A related reason that supports study of quantum robots and their
interactions with an environment is that they provide a {\it very small}
first step towards a quantum mechanical description of systems that are
aware of their environment, make decisions, are intelligent, and create
theories such as quantum mechanics \cite{Penrose,Stapp,Squires}. If quantum
mechanics is universal, then these systems must be described with quantum
mechanics to the maximum extent possible.

Another reason that supports study of quantum robots is that there is no 
limitation on the types of environments included. Environments studied so far,
such as oracles, data bases, and quantum registers, are special
types of environments.  These specific types of environments are discussed in 
the next section. Reasons are also given why the quantum robot must include
a quantum computer and cannot be simply a multistate head.

Section \ref{modelqr} summarizes a dynamical model of the
interactions of quantum robots with environments that has been
described elsewhere \cite{Ben1}.  The dynamics are described in
terms of tasks carried out by the quantum robot. Tasks are
defined as sequences of alternating computation and action
phases.  The  model describes task dynamics in terms of iteration of step
operators and Feynman \cite{Feynman} sums over phase paths.  

A simple example of a task, measuring the distance between a quantum
robot and a particle, is described in Section \ref{ex}. The
example is a generalization of the one described elsewhere in
that sums over different paths of states are included in the
phase path sum. A description of the task including the steps needed is 
given along with a representation of the task as a decision diagram.  
Accuracy conditions are also discussed. The complexity of even a simple
measurement, as a task for a quantum robot, and the relation to a possible 
Church Turing hypothesis \cite {Church,Deutsch} applicable to the carrying 
out of physical experiments \cite{Ben1} is noted in the last section.

\section{Oracles, Data Bases, Quantum Registers,and Heads} 

Oracles and data bases, as used in quantum oracle computation \cite{Bennett}
and in Grover's algorithm \cite{Grover}, and quantum registers, are special 
cases of environments considered here.  Oracles are special cases because the 
relevant properties are not supposed to be time dependent.  An oracle that 
answered yes to a question at one time and no to the same question later on 
would be regarded as defective. Also the answer to a question $Q$ at time 
$t$ should not be influenced by the asking of another question $P$ at an 
earlier time $t_{1}$ where both the yes and no answers to $P$ are ignored.

As environmental systems the states of both data bases and quantum registers
can be time dependent in the sense that data bases change as old data is
replaced or corrected or new data is added.  Also the states of quantum 
registers change with time as part of any quantum computation process.
However in both these cases the systems change in carefully specified ways.
In particular, neither of these systems are supposed to change
spontaneously in the absence of interaction with external systems.  An
environment of moving interacting systems whose states changed as a result
of the interactions or motion would not serve as a physical model of
 data bases or quantum registers.  In the latter case the requirement of no
spontaneous changes is emphasized by the need for quantum error correction 
codes and other stabilization methods to minimize this effect \cite{Laf,Zur}.

Environments considered here are not so limited.  They include 
interacting moving systems whose quantum states, $\Psi(t) =e^{-iHt}\Psi(0)$,
are evolving with time.  For these systems the  yes  answer to any question,
represented by a projection operator $Q$, has a time dependent
probability $\langle \Psi(t)|Q|\Psi(t)\rangle$. Also let $P$ and $Q$ be two 
projection operators corresponding to question $P$ asked at time $t_{1}$ and $Q$ 
asked at a later time $t$. The probability of a yes answer to $Q$ at $t$ is 
not in general equal to the probability of a yes answer to $Q$ at time $t$
given that $P$ was asked at time $t_{1}$ and the answer ignored. The latter 
probability is given by $Tr \rho(t)Q = Tr \rho(t_{1}) Q(t-t_{1})$ where 
$Q(t-t_{1}) =e^{iH(t-t_{1})}Qe^{-iH(t-t_{1})}$
and $\rho(t_{1})= PR_{\Psi(t_{1})}P + (1-P)R_{\Psi(t_{1})}(1-P)$ with
$R_{\Psi(t_{1})} =|\Psi(t_{1})\rangle \langle \Psi(t_{1})|$. This nonequality
holds in the case that $Q(t-t_{1})$ does not commute with $P$.  

Arguments that the quantum robot must include a quantum computer and cannot
be simply a multistate head are based on the number of degrees of freedom
in the head. If the head is a single degree of freedom that must be responsive
to at least $N$ different alternatives of environmental
information, it must be possible to distinguish between $N$
different internal states of the head. For large values of $N$, as would be
the case for a general purpose or universal quantum robot able to carry out
many tasks in many different environments \cite{Ben1},
this is physically unreasonable. For example if  the
head is a single spin system, it is very difficult to distinguish $N$
different spin projection states.

In this case, and in any case where the number of alternatives
that must be distinguished by the head is exponentially large,
(e.g. distinguishing all bit strings of length n) the only
reasonable approach is to allow the number of degrees of
freedom in the head to be polynomial in $\log N$.  But this is
equivalent to requiring that the head include a quantum computer.
That is it must be a quantum robot.  The same argument holds if
the head  has a small number ($>1$) of degrees of freedom.

\section{A model of Quantum Robots with Environments}
\label{modelqr}

Here the description of a specific model of quantum robots with
environments will be summarized.  Some details are given elsewhere
\cite{Ben1}.  A quantum robot consists of an on board quantum
computer, a finite state output system o and a control qubit c.
The dynamics of thse systems and their environmental interactions
can be described by tasks consisting of alternating computation
and action phases. The goal of each  computation phase is to
determine the following action by generating a new state of o.
Input consists of the former state of o, any stored memory, and
observations of the neighborhood state of the environment. During the 
following action phase the action, determined by the
state of o, is carried out. The states of all on board systems
remain unchanged.  Actions include motion of the quantum robot
and neighborhood changes of the environment state. The function of the control
qubit c is to turn on and off the two types of phases. The
computation [action] phase is inactive when c is in state $\vert
1\rangle [\vert 0\rangle ]$.  Each phase terminates by changing the state of c.

A unitary step operator $T=T_{a} +T_{c}$ which describes the
dynamical changes of the overall system during one time step is
associated with each task. (To keep things simple discrete space
and time are assumed.) If $\Psi(0)$ is the overall quantum robot
plus environment state at time $0$ the state after $n$ time steps
is given by $\Psi(n)=T^{n}\Psi(0)$.  The action and computation
phase step operators satisfy $T_{a}=TP^{c}_{1}$ and
$T_{c}=TP^{c}_{0}$ where the projection operators $P^{c}_{i}$
refer to the states of c.

Here $T_{c}$ is such that it may depend on but not change the quantum robot
location $\underline{x} =x,y,z$.  This can be expressed by a diagonality condition
\begin{equation}
T_{c} = \sum_{\underline{x}}P^{qr}_{\underline{x}} 
T_{c} P^{qr}_{\underline{x}}P^{c}_{0}. \label{Tcdiag}
\end{equation}
The requirement that $T_{a}$ depend on but not change the states of o is
given by a similar expression:
\begin{equation}
T_{a} =\sum_{l} P^{o}_{l}T_{a}P^{o}_{l}P^{c}_{1}. \label{Tadiag}
\end{equation}
The independence of $T_{a}$ from the quantum computer states $|b\rangle$ is
given by the commutativity of $T_{a}$ with  projection operators
$P^{qc}_{b}$.

Both $T_{c}$ and $T_{a}$ make local changes only in the environment state.
However those made by $T_{c}$ are limited to entanglements with on board 
system states and other changes that are a direct result of the observation
interaction.  Changes made by $T_{a}$ do not give such entanglements and are
not limited to observation interactions. Details on locality requirements
given in terms of neighborhoods of the quantum robot are given elsewhere
\cite{Ben1}.

The description of $T$ given so far applies to motionless noninteracting
environment systems for which the environment Hamiltonian $H_{E}=0$.  
Extension to moving interacting systems can be done by
replacing $T$ by another step operator ${\cal T} = T^{1/2}_{E}TT^{1/2}_{E} =
T^{1/2}_{E}T_{a}T^{1/2}_{E} + T^{1/2}_{E}T_{c}T^{1/2}_{E} = {\cal T}_{a} +
{\cal T}_{c}$. Here $T_{a}$ and $T_{c}$ are as defined above and $T_{E}
=e^{-iH_{E}\Delta}$ is 
a unitary step operator for the environment.  This replacement of $T$ by
$\cal T$ becomes exact in the limit $\Delta \rightarrow 0$ \cite{ReedSimon}. 

It is useful to express the dynamical development of the system
as a Feynman \cite{Feynman} sum over paths of computation and
action phases, i.e. a phase path sum. To this end consider the matrix 
element $\langle w,i\vert {\cal T}^{n}\vert w_{1},0\rangle$ which gives the 
transition amplitude for going from state $\vert w_{1},0\rangle$ to state
$\vert w,i\rangle$ in n steps. Here $|w\rangle$ denotes the state of all
systems except that of the control qubit.  One can use ${\cal T}^{n} =({\cal
T}(P^{c}_{0} +P^{c}_{1}))^{n}$ to obtain
\begin{equation}
\langle w,i\vert {\cal T}^{n}\vert w_{1},0\rangle = \sum_{t=1}^{n}
\sum_{\stackrel{ \scriptstyle{paths \; p\; \; of}} 
{length \;  t+1}}  \sum_{h_{1},\cdots
,h_{t} =1}^{\delta(\sum ,n)}   \langle p(t+1),i\vert
({\cal T}_{v_{t}})^{h_{t}}\vert p(t)\rangle  ,\cdots , 
\langle p(3)\vert ({\cal T}_{a})^{h_{2}}\vert p(2)\rangle  \langle
p(2)\vert ({\cal T}_{c})^{h_{1}}\vert p(1),0\rangle \label{pathsum}
\end{equation}
Each term in this large sum give the amplitude for finding $t$ alternating
phases in the first $n$ steps where the $j$th phase begins with all systems
(except for c) in state $|p(j)\rangle$ and ends after $h_{j}$ steps with all 
systems in state $|p(j+1)\rangle$. The upper limit on the $h$ sums shows the
restriction that the sum $h_{1}+,\cdots ,+h_{t} =n$. The initial and final
path states $|p(1)\rangle$ and $|p(t+1)\rangle$ are $|w_{1}\rangle$ and
$|w\rangle$.

Eq. \ref{pathsum} is shown for the case that the initial phase is a
computation phase, or c is in the initial state $|0\rangle$. A similar
equation holds in case the initial phase is an action phase. The alternation
of phases is expressed for this case by the subscripts $v_{j}$: if $j$ is
even $v_{j} =c$, if $j$ is odd $v_{j}=a$. Also the restrictions of Eqs.
\ref{Tadiag} and \ref{Tcdiag} on $T_{a}$ and $T_{c}$, which also hold for 
${\cal T}_{a}$ and ${\cal T}_{c}$,  show that if $j$ is even, $|p(j)\rangle$
and $|p(j+1)\rangle$ show the same robot position states.
If $j$ is odd, $|p(j)\rangle$ and $|p(j+1)\rangle$ show the
same quantum computer and o system states.

The equation shows clearly that for any $n$ the overall system state is a
linear sum over many phase path states of alternating computation
and action phases for the task represented by $\cal T$. For each value
of $t$ and $p$, the equation gives the amplitude for the phase
path $p$ containing $t-1$ completed phases and one, the $t$th, which may or
may not be complete. The $h$ sums give the dispersion
in the duration or number of time steps in each phase in $p$.

It follows from Eq. \ref{pathsum} that the overall system state $\Psi(n)$
can be expressed, for each initial state component, as an
exponentially growing (with $n$) tree of phase paths. Each node
in the tree corresponds to a state $|p(j)\rangle$.  The $t$ sum
shows that some branches of the tree have very few nodes and
include those having one node ($t=1$). 

\section{A Simple Example}
\label{ex}
\subsection{Task Description}

A simple example to illustrate the actions of a quantum robot
has an environment with a single motionless particle p on a 1D lattice (i.e.
$T_{E} =1$ and ${\cal T} =T$). The task is to measure the distance between the 
quantum robot and p by alternating quantum robot motion with local 
observations for p and counting the number of nonobservations of p  until p 
is found, (the search part).  In the return part, the quantum robot 
goes back the same number of steps and the task ends by entering the 
ballast part. This part is present to preserve the unitarity of $T$.

The overall goal of the task, as a condition on $T$ can be expressed as 
follows: Let $\phi =\sum_{y}c_{y}|y\rangle$ denote the state of p on the 
lattice with the quantum robot in position and internal memory state 
$|x,\underline{0}\rangle$.  The step operator $T$ for this task is to be such 
that iteration generates to good accuracy the well known entanglement
\begin{equation}
\phi |x,\underline{0}\rangle \longrightarrow \sum^{\prime}_{y}c_{y}
|y\rangle|x,\underline{y-x}\rangle \label{meas}
\end{equation}
over a limited range of $y$ values ($0\leq y-x<2^{N}$ --see below), denoted by 
the $\prime$ on the $\sum$.  Here $|\underline{y-x}\rangle$ is the permanent 
memory qubit string state corresponding to the lattice distance $y-x$ (number 
of sites) in one direction only between p and the quantum robot.  This equation 
is easily generalized to the case that the quantum robot is in a wave packet
$\theta_{qr} =\sum_{x}d_{x}|x\rangle$ of position states to give
\begin{equation}
\phi_{p}\theta_{qr}|\underline{0}\rangle \longrightarrow
\sum^{\prime}_{x,y}c_{y}d_{x}|y\rangle |x,\underline{y-x}\rangle. \label{gmeas}
\end{equation}

For this task the quantum computer contains two circular qubit quantum
registers, one with $N+2$ qubits and the other with $N+1$ qubits, and a head 
moving on the registers. Both registers accomodate numbers up to  $2^{N}-1$ 
with one ternary qubit in each in state $|2\rangle$ serving as an origin.  
The $N+2$ qubit register is a running memory for the search part counting 
(with o in state $\vert mr1\rangle$) and includes a sign qubit. The other 
serves to hold a permanent copy of the number on the running memory when p is 
located. Whenever p is found, a computation phase ends the search part by 
changing the o state to $\vert ml1\rangle$, carrying out the copying, and 
subtracting $1$ from the running memory to begin the return part. In this
part the computation phases subtract 1 from the running memory (no environment 
observations) until $-1$ is obtained.  The state of o is now changed 
to $|dn\rangle$ to begin the ballast part.  The
computation phases subtract 1 from the running memory until $-(2^{N}-1)$ is 
reached when the state of o becomes $|ml>\rangle$. The o state $\vert
mr>\rangle$ is reached if the particle is not located during the search part 
of the task (i.e. not located in at most $2^{N}-1$ iterations of the 
search action phase). For accurate measurements this phase is entered if 
$y-x<0$ or $y-x>2^{N}-1$.

During all action phases the quantum robot moves with the action determined
by the state of o.  No observations are carried out.  The o states
$|ml>\rangle ,\; |mr>\rangle$ correspond to nonterminating action phases as
final task parts.
The dynamics of the task can be given schematically by a decision diagram
which takes the properties of $T_{a}$ and $T_{c}$ into account.  This is
shown in Figure 1 with details given in the figure caption. The diagram is 
constructed so that it applies to the subtree emanating from
each node of the phase path tree described by Eq. \ref{pathsum}. That is, it
shows what happens in the subtree based on the overall system state
describing the node. This follows from the fact that there is no  diagram 
reference to where the quantum robot or p are on the lattice.
$x_{QR}=x_{p}?$ refers only the presence or absence of p at the location of
the quantum robot, wherever that is.  Also there are no definite duration
times associated with either action phases (circles) or with computation
phase components (square boxes).

In earlier work $T$ was required to be such that just one phase
path in the phase path sum of Eq. \ref{pathsum} contributed.  The
dispersion in phase path length ($t$ sum) and durations of phases
($h$ sums) was present.  Here this restriction will be retained
for the computation phase only. For $T_{c}$ the single path condition is 
expressed in Eq. \ref{pathsum} by requiring that if $j$ is even, then for each
input state $|p(j)\rangle$ to the $j/2$th computation phase,
there is a unique output phase path state $|p(j+1)\rangle$. Spreading or
quantum dispersion from $|p(j)\rangle$ to $|p(j+1)\rangle$, which  is
retained, is limited to that given by the $h_{j}$ sum (and the $t$ sum). 

Here the $T_{a}$ matrix elements
$\langle x^{\prime}, l,i|T_{a}|x,l,1\rangle$ are required to be local in the 
sense that their magnitude decreases  rapidly as the distance $|x^{\prime}-x|$ 
increases. The state of c is denoted by $i=0,1$. This is a
generalization of the earlier discussion of this example \cite{Ben1} in which 
only one phase path was included by requiring the $T_{a}$ matrix element to be $0$ 
unless $x{\prime} = x$ and $i=1$, or $x^{\prime}=x+1$ and $i=0$ for $|l \rangle =
|mr1\rangle$. For $|l \rangle = |ml1\rangle$ the second condition was
replaced by $x^{\prime}=x-1$ and $i=0$.

Phase path dispersion enters here in that the matrix elements $\langle
l,x^{\prime},i\vert T_{a}\vert l,x,1\rangle$ are $\neq 0$ for different
values of $x^{\prime}-x$. This generates many different action phase output
states for each input state with amplitudes determined by sums of products
of $T_{a}$ matrix elements over all paths {\it  within} the action phase. The
dependence of the matrix elements on the different final c states
$|i\rangle$ shows the action phase contributions to dispersion in
the number of phases in a path ($t$ sum) and duration in each
action phase ($h$ sum) in the phase path sum of Eq.
\ref{pathsum}.

\subsection{Accuracy}
It is clear from the task description that without further restrictions on
$T$, iteration of $T$ will describe a task and final overall system state that 
has no relation to the task goal as shown by Eq. \ref{meas}. To see this  let 
$\Psi(0)=\Theta (0)\phi_{p}$ represent an initial state with o and c in
state $|mr1,0\rangle$ and the quantum robot and p in the lattice position
state $|x\rangle \phi_{p}$ with $\phi_{p}=\sum_{y}c_{y}|y\rangle$. 
Other initial state values can be obtained from
Fig. 1.  The probability after $k$ steps that the search part of
the task is done and $\underline{n}$ is recorded on the permanent
memory is given by 
\begin{equation}
P_{k}(n) =\langle\Psi (k)\vert P^{st}_{\underline{n}}(1-P^{o}_{mr1})\vert \Psi 
(k)\rangle = \sum_{y}|c_{y}|^{2}P_{k}(n,y) \label{prob}
\end{equation}
where 
\begin{equation}
P_{k}(n,y)=\langle y,\Theta_{k}
(y)|P^{st}_{\underline{n}}(1-P^{o}_{mr1})|y,\Theta_{k}(y)\rangle. \label{proby}
\end{equation}
In these equations  $\Psi (k) =
T^{k}\Psi(0)=\sum_{y}c_{y}\Theta_{k}(y)|y\rangle$ where $\Theta_{k}(y)$ is
the state of the quantum robot after $k$ steps corresponding to p in state
$|y\rangle$. The projection operator $P^{st}_{\underline{n}}
=|\underline{n}\rangle \langle \underline{n}|$ where
$|\underline{n}\rangle$ is the permanent memory qubit string
state corresponding to the number $n$. The righthand equality of
Eq.\ref{prob}, and Eq. \ref{proby}, express the condition that p is
motionless and its state, except for possible entanglement, remains unchanged 
throughout the task.  

The probability $P_{k}(n,y)$ selects all phase paths in Eq. \ref{pathsum} 
containing $2n+1$ phases ($n$ action and $n+1$ computation phases) in the 
completed search part of the task. These paths have in common the fact that
in all but the last computation phase p was not at the location of the
quantum robot.   This corresponds in the search part of the phase path sum
to a limitation, for all but the last two phases, to states in which the
quantum robot and p are not at the same location. The sum over the output
states for the last ($n$th) action phase of the search part are limited to 
states in which the quantum robot and p are at the same location $y$.

The $k$ dependence of $P_{k}(n)$  enters through the requirement that  
paths in the phase path sum for $\Theta _{k}(y)|y\rangle$ contributing to 
Eq. \ref{proby} are those containing $n$ action
phases in a completed search part of the task within $k$ steps. The $k$
dependence enters through the $h$ sums of Eq. \ref{pathsum}, which express the
quantum dispersion in the durations of the different phases.  It depends
sensitively on the properties of $T$ and on the distance $y-x$. For 
sufficiently large $k$ the amplitudes of paths, that are still in search parts 
of the task with $<n+1$ completed computation phases, should be quite small.  
In this case if $T_{a}$ is reasonable (e.g. local, etc.) and $\phi$ is a wave
packet localized around some value $y_{0}$, then for values of
$y$ around $y_{0}$,  the infinite time limit $P_{\infty}(n,y)$ should exist 
and be sensibly equal to $P_{k}(n,y)$ for large $k$ if $0\leq y_{0}-x < 2^{N}$. 

It may also be the case that for large k the distribution of $P_{k}(n,y)$, as 
a function of $n$ may have a peak value and spread around the peak
value. However, without further restrictions on $T_{a}$, the peak value may
have no relation to the distance  between p and the quantum robot.
One way to remedy this is to require that the  matrix elements for $T_{a}$ 
have the form $\langle mr1, x^{\prime},i \vert T_{a}\vert 1, x, mr1\rangle
=  a_{i}e^{-\alpha (x^{\prime}-x-1+i)^{2}}$ where $a_{i}$ is an $i$
dependent coefficient.. In this case,
for large $\alpha$ and $k$, $P_{k}(n,y)$ should be peaked
around the value $n=y-x$ with a narrow dispersion (provided
$0\leq y-x<2^{N}$). In the limit $\alpha ,\; k = \infty$  the distance
measurement would be completely accurate with no dispersion in
the result. In this case $P_{\infty}(n,y) = \delta_{n,y-x}$ which agrees
with Eq. \ref{meas}.  The limit $k\rightarrow
\infty$ is needed because phase duration dispersion is present in the phase
path sum (the exponent is $0$ if either $x^{\prime} =x+1$ and $i=0$ 
or $x^{\prime} =x$ and $i=1$).

Generalization to the case in which Eq. \ref{gmeas} applies is
straightforward.  In this case the right hand of Eq. \ref{prob} is replaced
by $\sum_{x,x^{\prime},y}d^{*}_{x^{\prime}}d_{x}|c_{y}|^{2} 
P_{k}(n,x^{\prime},x,y)$. The sum is nondiagonal in $x$ and diagonal in $y$
because in this simple example, the quantum robot moves and p is motionless.
For large values of $\alpha ,\; k$ the values of
$P_{k}(n,x^{\prime},x,y)$ would be quite small for $ x^{\prime} \neq x$. In
the limit $\alpha =\infty$, $P_{\infty} (n,x^{\prime},x,y) =
\delta_{x^{\prime}-x}\delta_{n,y-x}$ which agrees with Eq. \ref{gmeas} for
$0\leq y-x<2^{N}$.

\section{Discussion}
Here the existence of a step operator $T_{c}$ that
connects each input path state to a unique output path state in each
computation phase was assumed implicitly. The existence of such a $T_{c}$
follows from the fact that there exists a classical Turing machine step
operator whose iterations describe a unique state path {\it within} each 
computation phase.   A quantum version can be defined that allows
spreading along the unique path for each computation phase and thereby
introduces the dispersion shown by the $h$ sums of Eq. \ref{pathsum}.
Generalization to $T_{c}$ that include sums over different phase
states, as was done here for $T_{a}$, is left to future work.  

The discussion above shows that the description of even a simple distance 
measurement is relatively complex if specific account is taken of all the steps 
needed (see Fig. 1) to generate entanglements of the form of Eq. \ref{meas}.  
This is based on the representation of numbers as 
states of quantum registers and the time taken in various pieces of 
the task (e.g. the $+1$ or $-1$ operations in the computation phases) and
other factors.  For the task in question the
activities, as shown in Figure 1, represent a task implementation that
suggests the existence, for the class of all physical experiments
implementable by quantum robots, of a hypothesis \cite{Ben1} similar to 
the Church Turing Hypothesis \cite{Church} 
for quantum computers \cite{Deutsch}.

In conclusion it should be stressed that, as an inanimate physical system,
the quantum robot knows nothing about counting, or where it is on the
lattice at any time, or that it is measuring anything at all. It is simply
behaving according to the dynamics specified by $T$.
\section*{Acknowledgements}
This work is supported by the U.S. Department of Energy, Nuclear 
Physics Division, under contract W-31-109-ENG-38.

Figure 1. Decision Diagram for the Example Task. Task motion is
shown by the arrows.  The round circles $mr1,\; mr>,\; ml1,\;
ml>$, and $dn$ denote action phases. The square boxes denote
memory system states (d = running memory and st = permanent
memory), questions, and addition ($d=d+1$) and subtraction
($d=d-1$) of $1$.  The boxes and arrows between successive
actions show activities of each computation phase. The left hand
column shows the task search part dynamics. The central column,
with horizontal arrows only, shows memory state changes when p is
found, and the righthand column shows the return part dynamics.
Ballast activities are shown separately at the bottom. The changes in o
system states denoting the end of task parts are not shown as they are
easily found from the diagram.

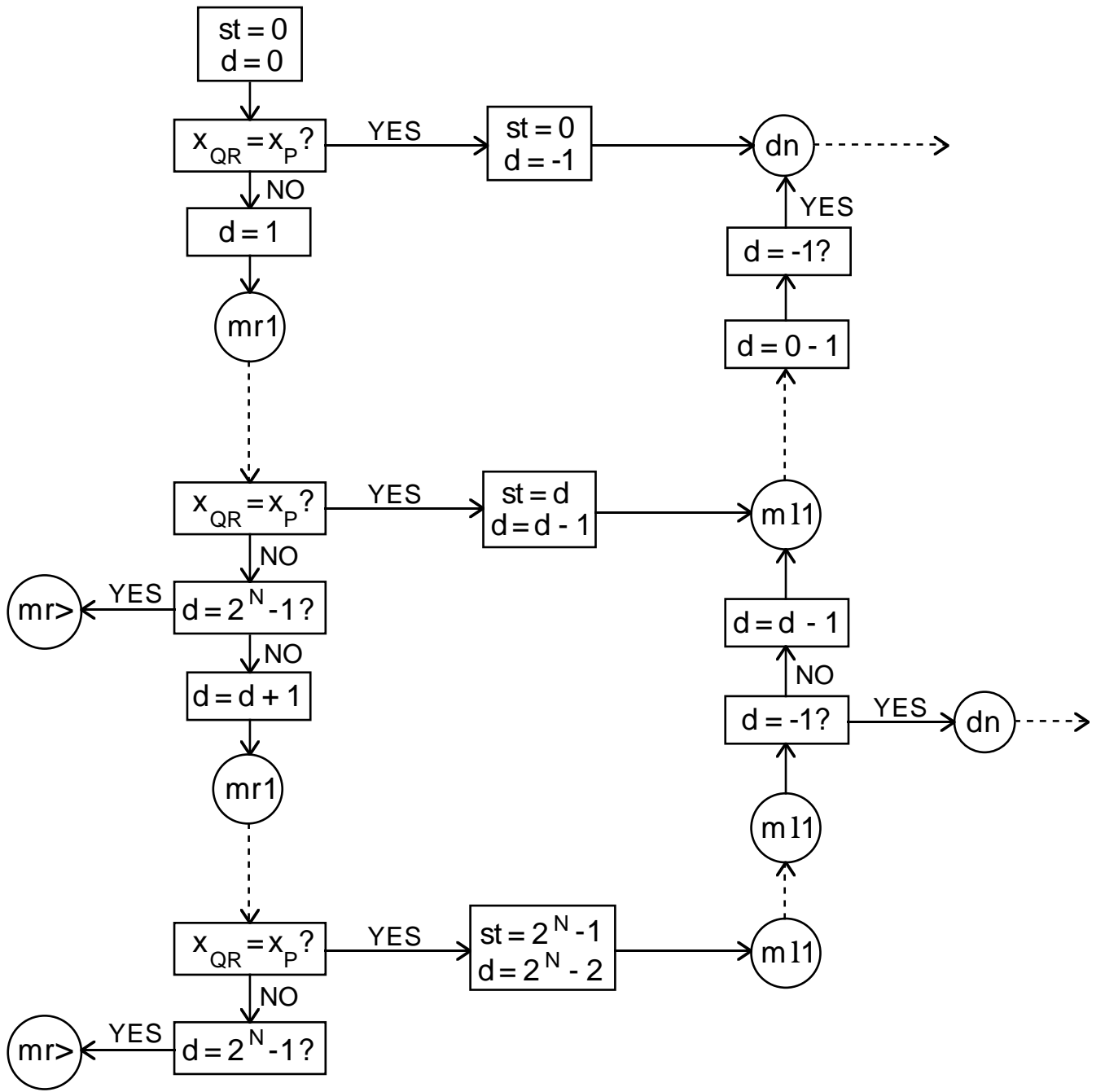
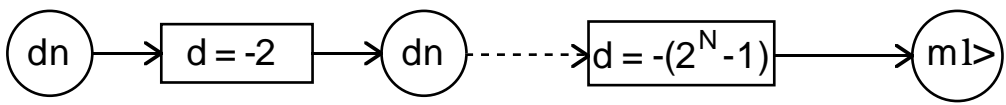
\end{document}